\def\tipo{1}

\def\av#1{\langle#1\rangle}          
\newcommand{\omitit}[1]{}

\ifnum \tipo = 2
\documentclass[twocolumn,showpacs,preprintnumbers,amsmath,amssymb,superscriptaddress]{revtex4}             
\def\figsize{8cm}
\def\subfigsize{4.2cm}

\else 
\documentclass[preprint,showpacs,preprintnumbers,amsmath,amssymb,superscriptaddress]{revtex4}
\def\figsize{12cm}
\def\subfigsize{6cm}

\fi

\usepackage{graphicx}
\usepackage{dcolumn}
\usepackage{bm}
\usepackage{subfigure}

\begin{document}

\preprint{APS/123-QED}

\title{Scale-Free Networks Emerging from Weighted Random Graphs}

\author{Tomer Kalisky}
\email{kaliskt@mail.biu.ac.il}
\affiliation{Minerva Center and Department of Physics, Bar-Ilan
  University, 52900 Ramat-Gan, Israel}
\author{Sameet Sreenivasan}
\affiliation{Center for Polymer Studies and Department of Physics
  Boston University, Boston, MA 02215, USA}
\author{Lidia A. Braunstein}
\affiliation{Center for Polymer Studies and Department of Physics
  Boston University, Boston, MA 02215, USA}
\affiliation{Departamento de F\'{\i}sica, Facultad de Ciencias Exactas
  y Naturales, Universidad Nacional de Mar del Plata, Funes 3350,
  $7600$ Mar del Plata, Argentina}
\author{Sergey V. Buldyrev}
\affiliation{Center for Polymer Studies and Department of Physics
  Boston University, Boston, MA 02215, USA}
\affiliation{Department of Physics, Yeshiva University, 500 West 185th
  Street, New York, NY 10033, USA}
\author{Shlomo Havlin}
\affiliation{Minerva Center and Department of
  Physics, Bar-Ilan University, 52900 Ramat-Gan, Israel}
\affiliation{Center for Polymer Studies and Department of Physics
  Boston University, Boston, MA 02215, USA}
\author{H. Eugene Stanley}
\affiliation{Center for Polymer Studies and Department of Physics
  Boston University, Boston, MA 02215, USA}


\date{\today} 

\begin{abstract}
  We study Erd\"{o}s-R\'enyi random graphs with random weights
  associated with each link. We generate a new ``Supernode network''
  by merging all nodes connected by links having weights below the
  percolation threshold (percolation clusters) into a single node. We
  show that this network is scale-free, i.e., the degree distribution
  is $P(k)\sim k^{-\lambda}$ with $\lambda=2.5$.
  Our results imply that the minimum spanning tree (MST) in random
  graphs is composed of percolation clusters, which are interconnected
  by a set of links that create a scale-free tree with $\lambda=2.5$.
  We show that optimization causes the percolation threshold to emerge
  spontaneously, thus creating naturally a scale-free ``supernode
  network.''
  We discuss the possibility that this phenomenon is related to the
  evolution of several real world scale-free networks.

\end{abstract}



\maketitle


\def\figureI{
  \begin{figure}
    \resizebox{\figsize}{!}{\includegraphics{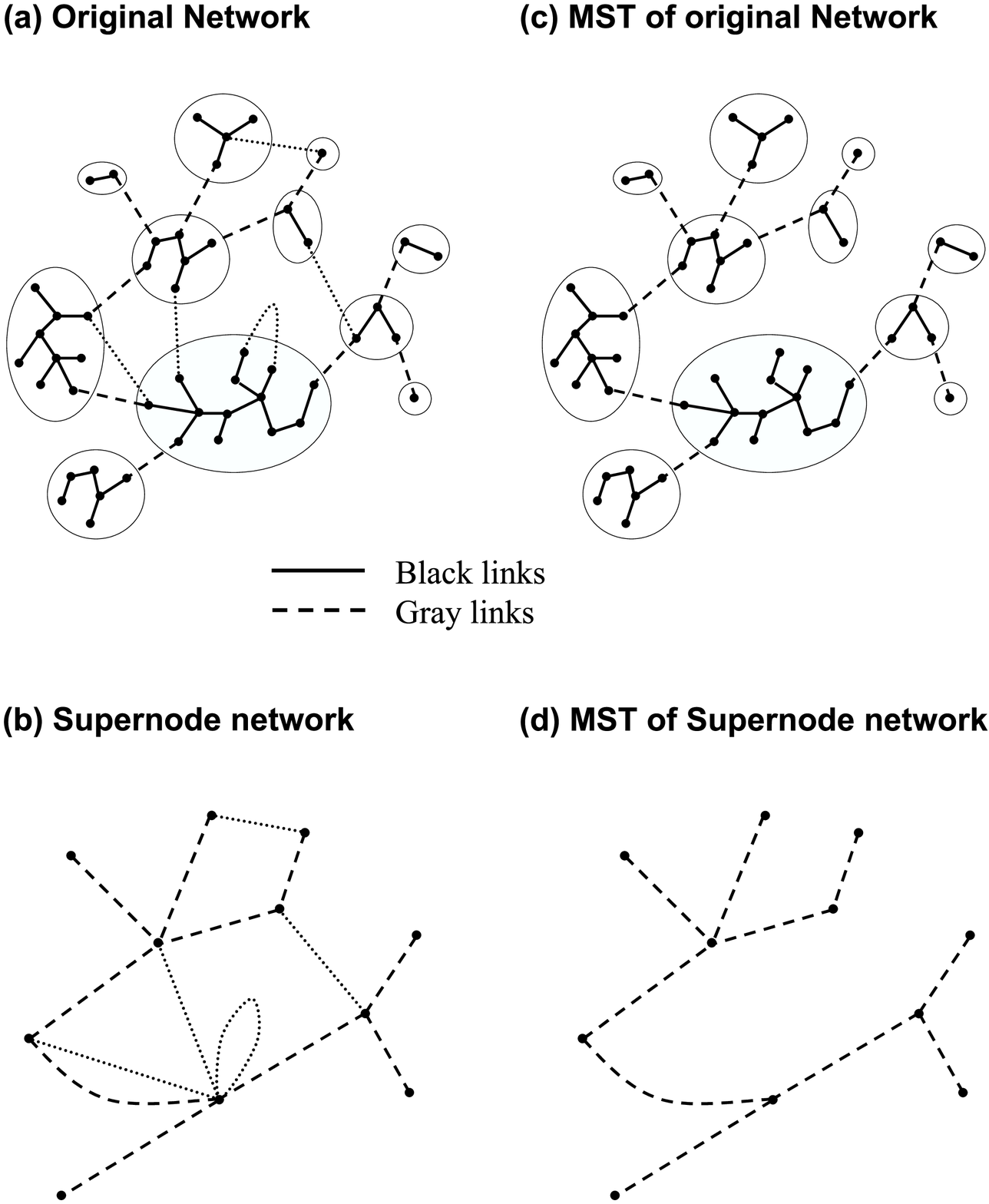}}
    \caption{\label{fig:gray_network_sketch} Sketch of the ``supernode
      network''. (a) The original ER network, partitioned into
      percolation clusters whose sizes $s$ are power-law distributed, with
      $n_s \sim s^{-\tau}$ where $\tau = 2.5$ for ER graphs. The ``black''
      links are the links with weights below $p_c$, the ``dotted''
      links are the links that are removed by the bombing algorithm,
      and the ``gray'' links are the links whose removal will
      disconnect the network (and therefore are not removed even
      though their weight is above $p_c$).
      (b) The ``supernode network'': the nodes are the clusters in the
      original network and the links are the links connecting nodes in
      different clusters (i.e., ``dotted'' and ``gray'' links). The
      supernode network is scale-free with $P(k)\sim k^{-\lambda}$ and
      $\lambda=2.5$. Notice the existence of self loops and of double
      connections between the same two supernodes.
      (c) The minimum spanning tree (MST), composed of black and gray
      links only.
      (d) The MST of the supernode network (``gray tree''), which is
      obtained by bombing the supernode network (thereby removing the
      ``dotted'' links), or equivalently, by merging the clusters in
      the MST to supernodes. The gray tree is scale-free, with
      $\lambda=2.5$.}
  \end{figure} }

\def\figureII{
  \begin{figure}
    \resizebox{\figsize}{!}{\includegraphics{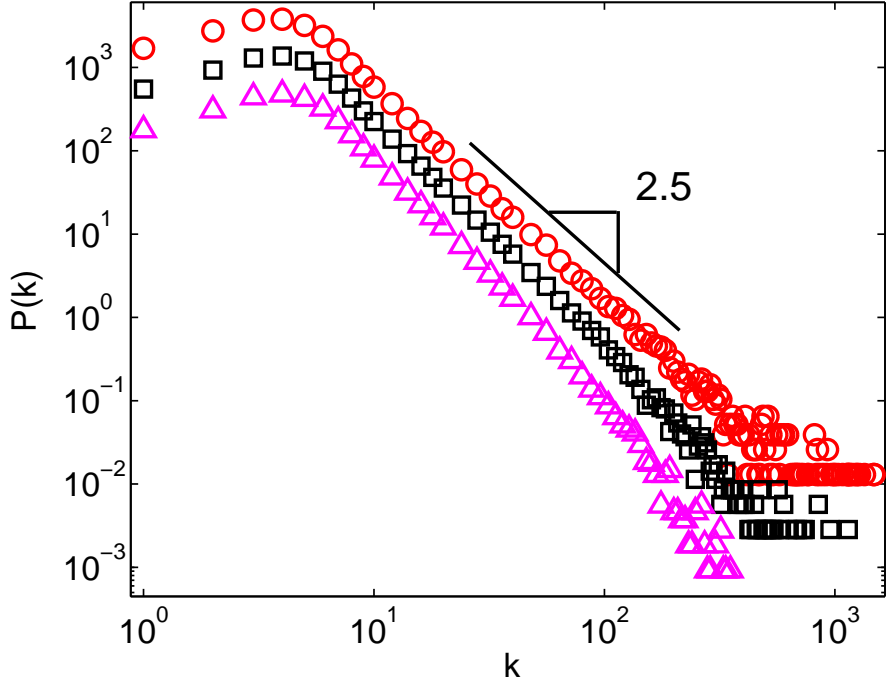} }
    \caption{\label{fig:graynet-degrees-layers} The degree distribution
    of the supernode network of Fig.~1(b), where the supernodes are the
    percolation clusters, and the links are the links with weights
    larger than $p_c$ ($\bigcirc$). The distribution exhibits a
    scale-free tail with $\lambda\approx 2.5$.  If we choose a threshold
    less than $p_c$, we obtain the same power law degree distribution
    with an exponential cutoff.  The different symbols represent
    slightly different threshold values: $p_c-0.03$ ($\Box$) and
    $p_c-0.05$ ($\bigtriangleup$).  The original ER network has
    $N=50,000$ and $\langle k \rangle =5$.  Note that for
    $k\approx\av{k}$ the degree distribution has a maximum.}
  \end{figure} }

\def\figureIII{
  \begin{figure} \resizebox{\subfigsize}{!}{
    \includegraphics{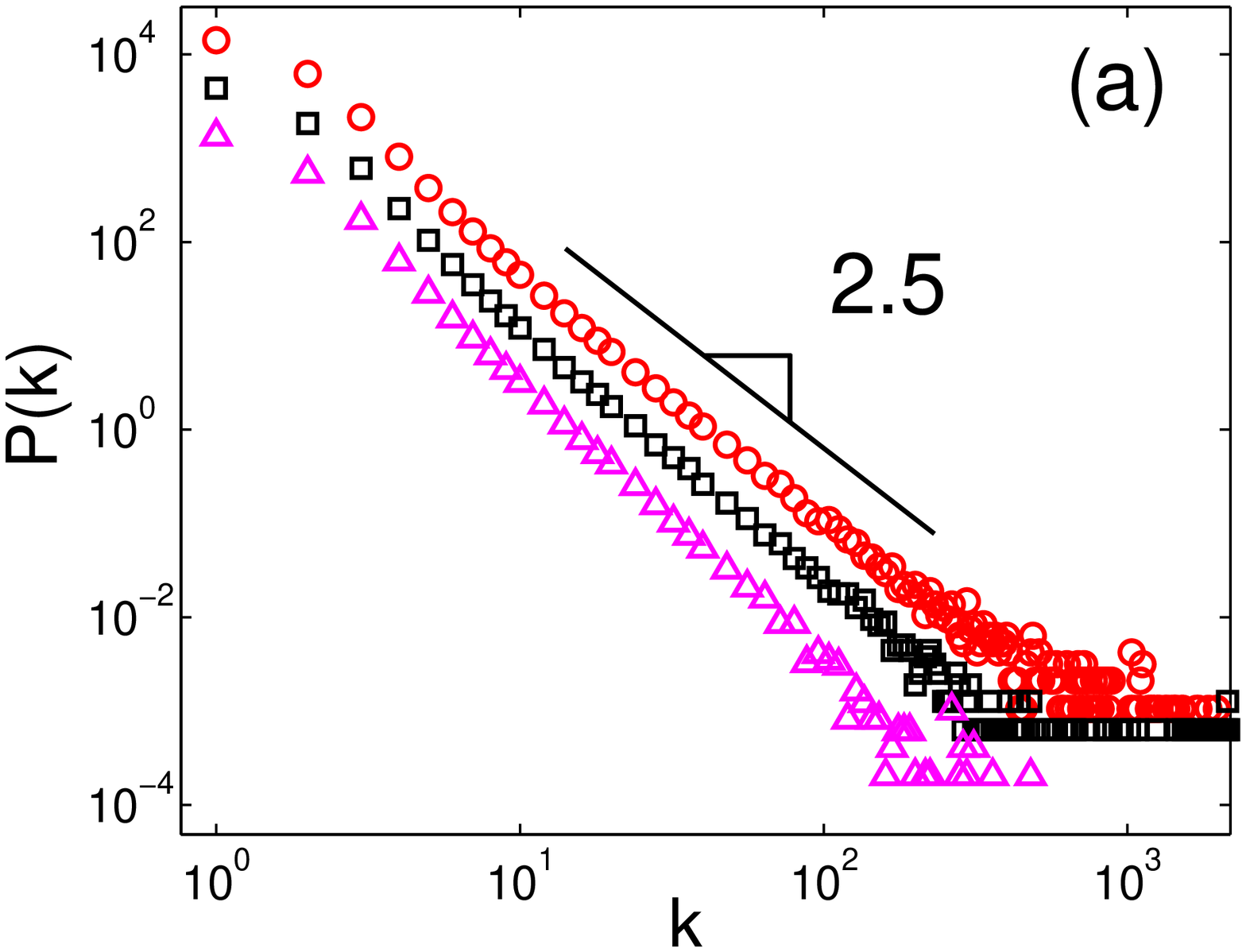} } \resizebox{\subfigsize}{!}{
    \includegraphics{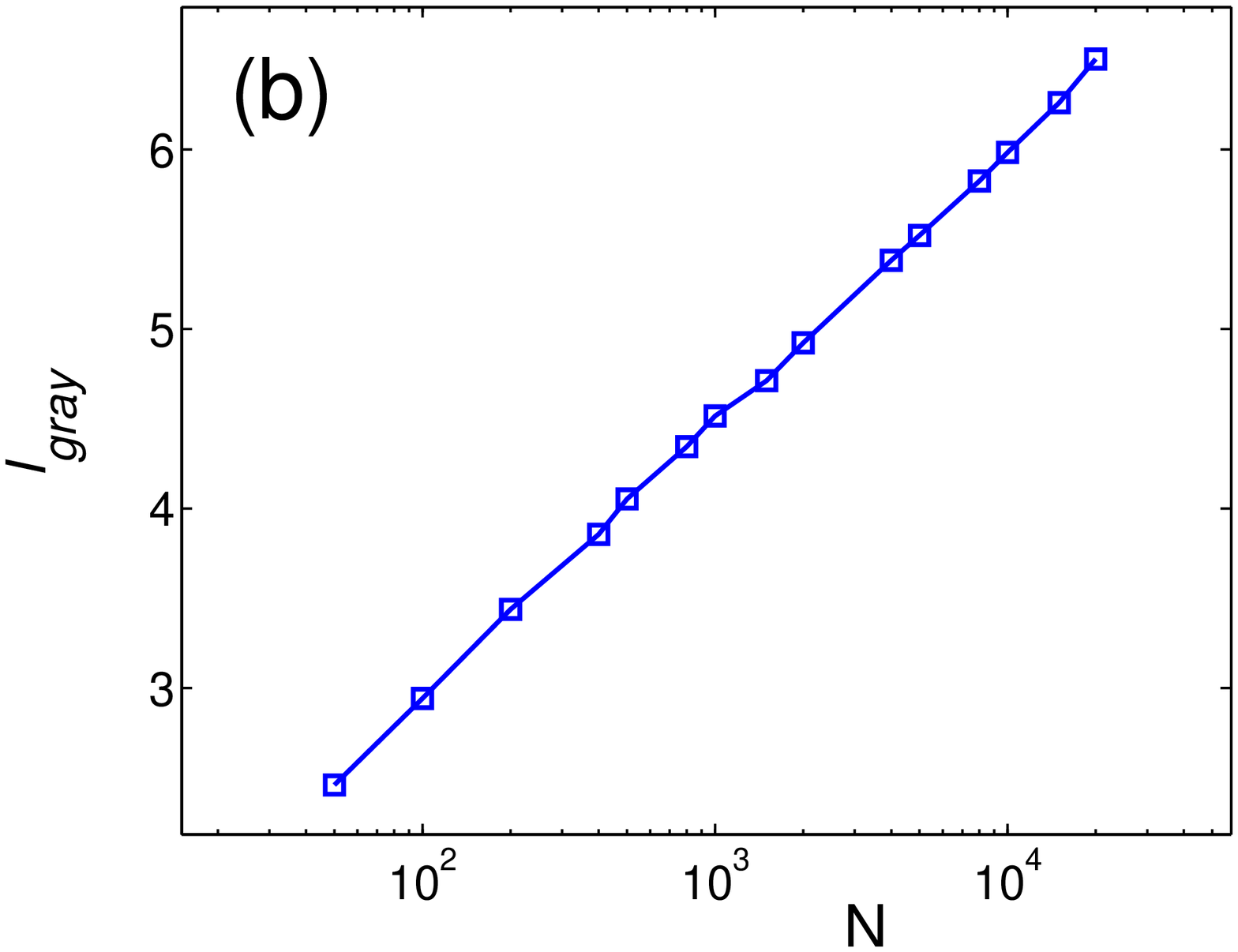} }
    \caption{\label{fig:tree_degrees_radius} (a) The degree distribution
      of the ``gray tree'' (the MST of the supernode network, shown in
      Fig.~\ref{fig:gray_network_sketch}(d)), in which the supernodes
      are percolation clusters and the links are the gray links.
      Different symbols represent different threshold values: $p_c$
      ($\bigcirc$), $p_c+0.01$ ($\Box$) and $p_c+0.02$
      ($\bigtriangleup$). The distribution exhibits a scale-free tail
      with $\lambda\approx 2.5$, and is relatively insensitive to
      changes in $p_c$.
    (b) The average path length $\ell_{\rm gray}$ on a the gray tree as
      a function of original network size. It is seen that $\ell_{\rm
      gray}\sim\log N_{\rm sn}\sim\log N$.  \omitit{The same behavior can be seen for
      MST's on random scale-free graphs with $\lambda=2.5$ if we allow
      for double connections.}}  
  \end{figure} }

\def\figureIV{
  \begin{figure}
    \resizebox{\figsize}{!}{\includegraphics{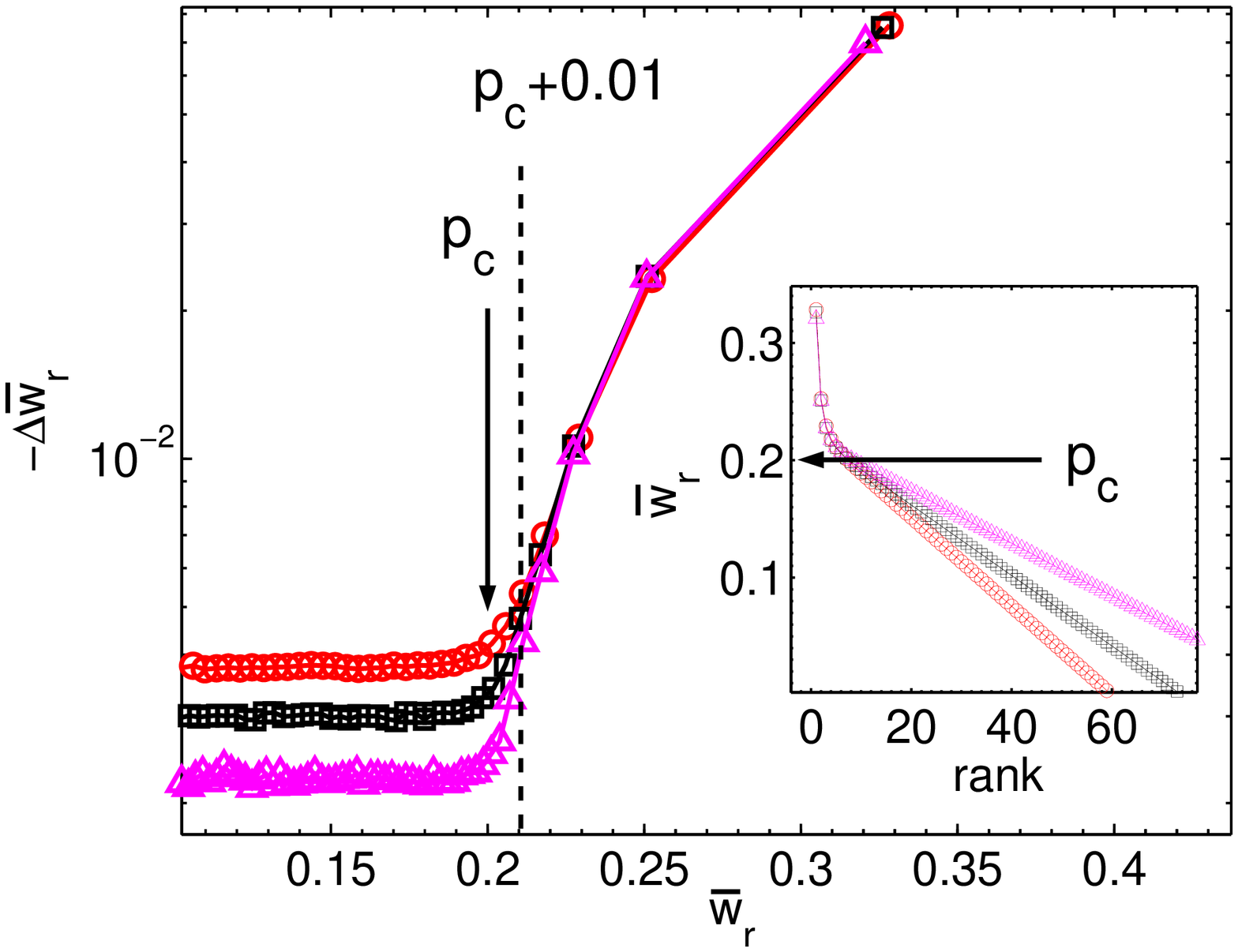}}
    \caption{\label{fig:weights_optimal_path} The inset shows, for an ER
      graph with $\av{k}=5$, the average weights $\bar w_r$ along the
      optimal paths, sorted according to their rank.  The main figure
      shows $\Delta\bar w_r\equiv \bar w_{r+1}-\bar w_r$, where $\bar
      w_r$ is the mean weight for rank $r$, vs. the weights along the
      optimal path. Different symbols represent different system
      sizes: $N = 8000$ ($\bigcirc$), $N = 16,000$ ($\Box$) and $N =
      32,000$ ($\bigtriangleup$). Below $p_c=0.2$, $\Delta\bar w_r$
      decreases for increasing $N$, while weights $\bar w_r$ well
      above $p_c$ do not change with N.}
\end{figure} }

Scale-free topology is very common in natural and man-made networks.
Examples vary from social contacts
between humans to technological networks such as the World Wide Web or
the Internet
\cite{Barabasi-Albert-2002:statistical-mechanics,Dorogovtsev-Mendes-2003:From_Biological_Nets_to_the_Internet,vespignani-pastor-satorras-2004:evolution_and_structure}.
Scale free (SF) networks are characterized by a power law distribution
of connectivities $P(k) \sim k^{-\lambda}$, where $k$ is the degree of
a node and the exponent $\lambda$ controls the broadness of the distribution. Many
networks are observed to have values of $\lambda$
around $2.5$. For values of $\lambda < 3$ the second moment of the
distribution $\langle k^2 \rangle$ diverges, leading to several
anomalous
properties~\cite{multiple_papers-cohen-havlin-2000-2003:resilience_ultra}.

In many real world networks there is a ``cost'' or a ``weight''
associated with each link, and the larger the weight on a link, the
harder it is to traverse this link. In this case, the network is
called ``weighted''~\cite{vespignani-2004:weighted_evolving_networks}.
Examples can be found in communication and computer networks, where
the weights represent the bandwidth or delay time, in protein networks
where the weights can be defined by the strength of interaction
between
proteins~\cite{uets-2000:interacting_proteins,rao-caflisch-2004:protein_folding}
or their structural
similarity~\cite{Dokholian-Shakhnovich-2002:Expanding}, and in
sociology where the weights can be chosen to represent the strength of
a
relationship~\cite{Barabasi-2002:Linked,newman-2001:collaboration_shortest_II}.

In this Letter we introduce a simple process that generates random
scale-free networks with $\lambda=2.5$ from weighted Erd\"{o}s-R\'enyi
graphs~\cite{Erdos-Renyi-1960:Evolution}.  We further show that the
minimum spanning tree (MST) on an Erd\"{o}s-R\'enyi graph is related
to this network, and is composed of percolation clusters, which we
regard as ``super nodes'', interconnected by a scale-free tree.
We will see that due to optimization this scale-free tree is dominated
by links having high weights --- significantly higher than the
percolation threshold $p_c$. Hence, the MST naturally distinguishes
between links below and above the percolation threshold, leading to a
scale-free ``supernode network''. Our results may explain the origin
of scale-free degree distribution in some real world networks.


Consider an Erd\"{o}s-R\'enyi (ER) graph with $N$ nodes and an average
degree $\av{k}$, thus having a total of $N \langle k \rangle /2$
links. To each link we assign a weight chosen randomly and uniformly
from the range $[0,1]$. We define black links to be those links with
weights below a threshold $p_c=1/\langle k\rangle$
\cite{Erdos-Renyi-1960:Evolution}. Two nodes belong to the same
cluster if they are connected by black links [Fig.~1(a)]. From
percolation theory
\cite{Bunde-Havlin-1996:Fractals-and-Disordered-Systems} follows that
the number of clusters of $s$ nodes scales as a power law, $n_s \sim
s^{-\tau}$, with $\tau=2.5$ for ER networks \cite{text1}.
Next, we merge all nodes inside each cluster into a single
``supernode'' \cite{text2}.
We define a new ``supernode network'' [Fig.~1(b)] of $N_{\rm sn}$
supernodes
\cite{Sreenivasan-Kalisky-Braunstein-Buldyrev-Havlin-Stanley-2003:Effect}.
The links between two supernodes [see
Figs.~\ref{fig:gray_network_sketch}(a) and
\ref{fig:gray_network_sketch}(b)] have weights {\em larger\/} than
$p_c$.

\ifnum\tipo=2
\figureI
\fi

The degree distribution $P(k)$ of the supernode network can be
obtained as follows. Every node in a supernode has the same (finite)
probability to be connected to a node outside the supernode. Thus, we
assume that the degree $k$ of each supernode is proportional to the
cluster size $s$, which obeys $n_s\sim s^{-2.5}$.  Hence $P(k)\sim
k^{-\lambda}$, with $\lambda=2.5$, as supported by simulations shown
in Fig.~\ref{fig:graynet-degrees-layers}.


\ifnum\tipo=2
\figureII
\fi


We next show that the minimum spanning tree (MST) on an ER graph is
related to the supernode network, and therefore also exhibits
scale-free properties. The MST on a weighted graph is a tree that
reaches all nodes of the graph and for which the sum of the weights of
all the links (total weight) is minimal. Also, each path between two
sites on the MST is the optimal path in the ``strong disorder'' limit
\cite{Cieplak94,Dobrin-Dexbury-1960:Minimum_Spanning_Trees}, meaning
that along this path the maximum {\em barrier} (weight) is the
smallest possible
\cite{Dobrin-Dexbury-1960:Minimum_Spanning_Trees,%
Braunstein-Buldyrev-Cohen-Havlin-Stanley-2003:Optimal,
Sreenivasan-Kalisky-Braunstein-Buldyrev-Havlin-Stanley-2003:Effect}.

Standard algorithms for finding the MST
\cite{Cormen-2001:Introduction} are Prim's algorithm, which resembles
invasion percolation, and Kruskal's algorithm, which resembles
percolation.
An equivalent algorithm to find the MST is the ``bombing algorithm''
\cite{Dobrin-Dexbury-1960:Minimum_Spanning_Trees,Braunstein-Buldyrev-Cohen-Havlin-Stanley-2003:Optimal}.
We start with the full ER network and remove links in order of
descending weights.  If the removal of a link disconnects the graph,
we restore the link and mark it
``gray''~\cite{Ioselevich-Lyubshin-2004:Phase}; otherwise the link
[shown dotted in Fig.~1(a)] is removed. The algorithm ends and an MST
is obtained when no more links can be removed without disconnecting
the graph.

In the bombing algorithm, only links that close a loop can be removed.
Because at criticality loops are negligible
\cite{Erdos-Renyi-1960:Evolution,Bunde-Havlin-1996:Fractals-and-Disordered-Systems}
for ER networks ($d\to\infty$), bombing does not modify the
percolation clusters --- where the links have weights below $p_c$.
Thus, bombing modifies only links {\em outside} the clusters, so
actually it is only the links of the {\em supernode network} that are
bombed.  Hence the MST resulting from bombing is composed of
percolation clusters connected by gray links
[Fig.~\ref{fig:gray_network_sketch}(c)].

From the MST of Fig.~\ref{fig:gray_network_sketch}(c) we now generate
a new tree, the MST of the supernode network, which we call the ``gray
tree'', whose nodes are the supernodes and whose links are the gray
links connecting them [see Fig.~\ref{fig:gray_network_sketch}(d)].
Note that bombing the original ER network to obtain the MST of
Fig.~\ref{fig:gray_network_sketch}(c) is equivalent to bombing the
supernode network of Fig.~\ref{fig:gray_network_sketch}(b) to obtain
the gray tree, because the links inside the clusters are not bombed.
%
We find [Fig~\ref{fig:tree_degrees_radius}(a)] that the gray tree has
also a scale-free degree distribution $P(k)$, with $\lambda=2.5$---the
same as the supernode network \cite{text3}.
We also find [Fig.~\ref{fig:tree_degrees_radius}(b)] the average path
length $\ell_{\rm gray}$ scales as $\ell_{\rm gray} \sim \log N_{\rm
  sn} \sim \log N$
\cite{Sreenivasan-Kalisky-Braunstein-Buldyrev-Havlin-Stanley-2003:Effect,text4}.
Note that even though the gray tree is scale-free, it is not
ultra-small~\cite{multiple_papers-cohen-havlin-2000-2003:resilience_ultra},
since the length does not scale as $\log\log N$.

\ifnum\tipo=2
\figureIII
\fi


Next we show that our optimization of the MST, which leads to the gray
tree, yields a significant separation between the weights of the links
inside the supernodes and the links connecting the supernodes.
We consider each pair of nodes in the original MST of $N$ nodes
[Fig.~1(c)] and calculate the typical path length $\ell_{\rm typ}$,
which is the most probable path length on the MST. For each path of
length $\ell_{\rm typ}$ we rank the weights on its links in descending
order. For the largest weights (``rank 1 links''), we calculate the
average weight $\bar w_{r=1}$ over all paths. Similarly, for the next
largest weights (``rank 2 links'') we find the average $\bar w_{r=2}$
over all paths, and so on up to $r=\ell_{\rm typ}$. The inset in
Fig.~\ref{fig:weights_optimal_path} shows $\bar w_r$ as a function of
rank $r$ for three different network sizes $N=8000$, $16000$, and
$32000$.  In Fig.~\ref{fig:weights_optimal_path} we plot the
difference in consecutive average weights, $\Delta\bar w_r\equiv\bar
w_r-\bar w_{r-1}$ as a function of $\bar w_r$.
We see that weights below $p_c$ (black links inside the supernodes)
are uniformly distributed and approach one another as $N$ increases.
As opposed to this, weights above $p_c$ (``gray links'') are {\it
  not\/} uniformly distributed, due to the bombing algorithm, and are
independent of $N$.
The latter links with the highest weights can be associated with gray
links from very small clusters [Figs.~1(a) and 1(c)]. These links
almost cannot be bombed due to limited number of exits from small
clusters, and therefore do not change with $N$. Moreover, because of
the abundance of small clusters ($n_s \sim s^{-\tau}$), large clusters
are connected mostly to small clusters (through links with relatively
large weights).

We thereby obtain a scale-free network with $\lambda=2.5$, which is
not very sensitive to the precise value of the threshold used for
defining the supernodes.
For example, the scale-free degree distribution shown in
Fig.~\ref{fig:tree_degrees_radius}(a) for a threshold of $p_c+0.01$
corresponds to having only four largest weights on the optimal paths
[see Fig.~\ref{fig:weights_optimal_path}]. This means that mainly very
small clusters, connected with high-weight links to large clusters,
dominate the scale-free distribution $P(k)$ of the MST of the
supernode network (gray tree).
Hence, the optimization process on an ER graph causes a significant
separation between links below and above $p_c$ to emerge {\it
spontaneously\/} in the system, and by merging nodes connected with
links of low weights, a scale-free network can arise.
\omitit{Hence, the optimization process on an ER graph causes the
  critical threshold $p_c$ to emerge spontaneously in the system, and
  by merging nodes connected with links of low weight, a scale-free
  network evolves.}

The process described above may be related to the evolution of some
real world networks. Consider a homogeneous network with many
components whose average degree $\langle k\rangle$ is well defined.
Suppose that the links between the components have different weights,
and that some optimization process separates the network into nodes
which are well connected (i.e., connected by links with low weights)
and nodes connected by links having much higher weights. If the
well-connected components merge into a single node, this results in a
new heterogeneous supernode network with components that vary in size,
and thus in number of outgoing connections.
\omitit{and that some {\em optimization} process eventually unites
  components that are strongly linked.  This results in a new
  heterogeneous network with components that vary in size, and thus in
  number of outgoing connections.}

An example of a real world network whose evolution may be related to
this model is the protein folding network, which was found to be
scale-free with $\lambda \approx 2.3$
\cite{rao-caflisch-2004:protein_folding}. The nodes are the possible
physical configurations of the system and the links between them
describe the possible transitions between the different
configurations. We assume that this network is {\em optimal\/} because
the system chooses the path with the smallest energy {\em barrier\/}
from all possible trajectories in phase space. It is possible that the
scale-free distribution evolves through a similar procedure as
described above for random graphs: adjacent configurations with close
energies (nodes in the same cluster) cannot be distinguished and are
regarded as a single supernode, while configurations (clusters) with
high barriers between them belong to different supernodes.

A second example is computer networks. Strongly interacting computers
(such as computers belonging to the same university) are likely to
converge into a single domain, and thus domains with various sizes and
connectivities are formed.  This network might be also optimal,
because packets destined to an external domain are presumably routed
through the router which has the best connection to the target domain.

\ifnum\tipo=2
\figureIV
\fi



To summarize, we have seen that any weighted random network hides an
inherent scale-free ``supernode network'' \cite{text5}.
We showed that the minimum spanning tree, generated by the bombing
algorithm, is composed of percolation clusters connected by a scale-free
tree of ``gray'' links. Most of the gray links connect small clusters to
large ones, thus having weights well above the percolation threshold
that do not change with the original size of the network.
Thus the optimization in the process of building the MST distinguishes
between links with weights below and above the threshold, leading to a
spontaneous emergence of a scale-free ``supernode network''. 
We raise the possibility that in some real world networks, nodes
connected well merge into one single node, and through a natural
optimization a scale-free network emerges.

\omitit{To summarize, the Minimum Spanning Tree is composed of
  clusters of links with weight below the critical probability $p_c$,
  which are connected by a scale-free tree of links above $p_c$. The
  average path length of this tree scales as $\log N$.}

\section*{Acknowledgments}
We thank the ONR, ONR-Global, the Israel Science Foundation and the
Israeli Center for Complexity Science for financial support. We thank
R. Cohen, E.~Perlsman, A.~Rosenfeld, Z.~Toroczkai, S.~Galam,
D.~Stauffer, P.~L.~Krapivsky and O.~Riordan for useful
discussions.

\appendix


\begin{thebibliography}{24}
\bibitem{text1} ER networks can be regarded as having an infinite
  dimension since space does not play any role.
\bibitem{text2} This is done by the following algorithm: we follow all
  links of the network. If there exists a link between any two nodes
  with weight smaller than $p_c = 1/\av{k}$, we identify the nodes as
  belonging to the same cluster.
\bibitem{text5} Similar results can also be obtained for graphs
  embedded in two or three dimensions, with different power law
  exponents.
\end{thebibliography}

  
  

\ifnum\tipo=1
\figureI
\figureII
\figureIII
\figureIV
\fi

\end{document}